# Four new active galaxies with steep soft X-ray spectra*

J. Greiner[1], R. Danner[1,2], N. Bade[3], G.A. Richter[4], P. Kroll[4], S. Komossa[1]

[1] Max-Planck-Institut für Extraterrestrische Physik, 85740 Garching, Germany
[2] Palomar Observatory, California Institute of Technology, Pasadena, CA 91125, U.S.A
[3] Sternwarte Hamburg, 21029 Hamburg, Gojenbergsweg 112, Germany
[4] Sternwarte Sonneberg, 96515 Sonneberg, Germany

**Abstract.** We have discovered four AGN in the *ROSAT* all-sky-survey data with very steep X-ray spectra. We apply several models to these X-ray spectra with emphasis on warm absorber models which give an adequate description of the data. We report on the follow-up optical and radio observations which allow the identification of three of these objects as Narrow Line Seyfert 1 galaxies, and the fourth as BL Lac object. We have measured small–FWHM H$\beta$ lines, strong FeII emission and weak [OIII] emission in the three Narrow Line Seyfert 1 galaxies, in line with known correlations with respect to the steepness of the X-ray spectra.

We have discovered strong optical variability in the BL Lac object and two of the Seyfert galaxies using photographic plates of the Sonneberg Observatory field patrol.

We finally discuss the statistical implications of our search algorithm on the expected number density of soft X-ray selected AGN and conclude that up to 30% of X-ray selected AGN might have supersoft X-ray spectra.

**Key words:** X-ray sources – AGN – soft X-ray excess

## 1. Introduction

Observations with the HEAO-1, *Einstein*, *EXOSAT* and *Ginga* satellites have shown that the X-ray spectra of active galactic nuclei (AGN) above a few keV are well described by a power law with a photon index of about –1.5 for radio-loud and about –1.9 for radio-quiet quasars. A soft X-ray excess below $\approx$1 keV is a common feature in the X-ray spectra of AGN. This excess is often related to the optical/UV big blue bump which dominates the spectra of most radio-quiet AGN. In some cases the excess at X-rays can be modeled as a very steep and soft component which is consistent with the Wien tail of a hot thermal component.

A systematic correlation of *ROSAT* all-sky-survey X-ray sources with known AGN has resulted in 102 sources with more than 80 counts suited for an estimation of their spectral parameter, in particular their hardness ratios (Schartel 1994). AGN in the radio-quiet subsample of this *ROSAT*-quasar sample show significantly steeper spectra (photon index $\Gamma = -2.53\pm0.04$) than those of the radio-loud subsample ($\Gamma = -2.27\pm0.07$).

Further detailed studies of selected AGN have revealed some objects with extremely steep soft X-ray spectra including IRAS 13324-3809 (Boller et al. 1993), Ark 564 (Brandt et al. 1994), the high redshift object E1346+266 (Puchnarewicz et al. 1994), IC 3599 (Brandt et al. 1995, Grupe et al. 1995a), and WPVS007 (Grupe et al. 1995b). A variety of different models has been proposed to describe the emission of these objects, but no generally accepted explanation has emerged yet. Reprocessing and free-free emission have been suggested already earlier, and recently accretion disk models with various modifications have become popular.

Optical properties of a large sample of *Einstein* ultrasoft AGN have been studied by Puchnarewicz et al. (1992). They found that a major part of these soft X-ray selected AGN turn out to be Narrow Line Seyfert 1 galaxies with narrow (FWHM <2000 km/s) H$\beta$ lines (Osterbrock and Pogge 1985).

Here we report the discovery of four very soft X-ray AGN which were found searching *ROSAT* all-sky-survey data. We present the optical identifications (section 2.2), details of the optical spectroscopy (section 2.3), describe the discovery of optical variability of three out of the four objects (section 2.4), give details on the X-ray spectra and the resulting parameters of the model fitting (section 2.5), present the survey X-ray lightcurves and derive X-ray luminosities (section 2.6), report the radio observations of three of the four objects (section 2.7), and finally discuss our observational and fitting results and the statistical implications of our search (section 3).

**Table 1.** New soft AGN in the Coma field

| ROSAT Name | No. of counts | Hardness ratio HR1 | Variable designation | Position of the optical counterpart (2000.0) | D |
|---|---|---|---|---|---|
| RX J1239.3+2431 | 120 | −0.75±0.07 | S 10940 | $12^h39^m18\overset{s}{.}8$ +24°31′44″ (± 1″) | 8″ |
| RX J1257.5+2412 | 455 | −0.09±0.05 | S 10941 | $12^h57^m31\overset{s}{.}9$ +24°12′40″ (± 1″)[1] | 8″ |
| RX J1225.7+2055 | 152 | −0.66±0.07 | S 10942 | $12^h25^m41\overset{s}{.}9$ +20°55′04″ (± 1″) | 5″ |
| RX J1250.2+1923 | 83 | −0.66±0.10 | – | $12^h50^m15\overset{s}{.}0$ +19°23′50″ (± 1″) | 3″ |

[1] Here, the radio position is given since the optical position is not accurate due to blending (see section 2.7).

## 2. Observational results

### 2.1. Selection criteria

In a study aimed at a statistical comparison of optically variable sources at different galactic latitudes we have examined ROSAT data in a 100 square degree field centered around 26 Com (for first results on flare stars see Richter, Bräuer & Greiner (1995) and on cataclysmic variables see Richter & Greiner (1995a, b)). The Coma field was scanned during the ROSAT all-sky-survey in December 1990 for a mean total observing time of 470 sec (Tab. 6). Using a maximum likelihood method we detected 238 X-ray sources in the above 10×10 degree field with a likelihood larger than 10. These X-ray sources were identified using (1) the objective prism spectra taken with the Hamburg Schmidt telescope on Calar Alto, (2) including the positional correlation with the X-ray positions which are accurate to typically less than 30″ and (3) the X-ray to optical intensity ratio for known populations. In the Hamburg objective prism survey (Hagen et al. 1995) spectra are taken in the 3400–5400 Å range with a dispersion of 1390 Å/mm down to 17–18th mag covering the whole northern hemisphere except the galactic plane (| b |>20°).

For the present purpose of looking for soft AGN we have applied the following three selection criteria:
1. The hardness ratio HR1 (defined as the number of counts in the PSPC channels (52–201)-(11–41) vs. those in channels (11–41)+(52–201)) plus its error is lower (i.e. softer) than −0.5.
2. The total number of counts collected during the all-sky-survey with the PSPC is greater than 80.
3. The object classification using the spectra of the objective prism plates indicates an extragalactic object, i.e. classifications of "QSO", "EBL-WK" (weak blue object with emission lines) or "Blue Gal." were chosen (see Bade et al. 1995 for details on the object classification).

This selection yielded three of the four sources listed in Tab. 1. In addition, we included the X-ray brightest non-stellar object (RX J1257.5+2412) in our field because of its soft spectrum (which does not fit the hardness ratio criterion, however). This source was previously detected in X-rays during an *Einstein* slew and then designated 1ES 1255+244 (Elvis et al. 1992).

All objects are new identifications. In Tab. 1 we give the ROSAT name (column 1), the total number of counts collected during the ROSAT all-sky-survey observation (2), the hardness ratio HR1 with error (3), the Sonneberg variable designation (4, see paragraph 2.4. for more details), the position of the optical counterpart (5), and the distance D between X-ray and optical position (6).

### 2.2. Optical identification

Inspection of the X-ray source positions on the Palomar Observatory Sky Survey revealed only one optical counterpart candidate each inside the $2\sigma$ error circle for RX J1225.7+2055 and RX J1250.2+1923, whereas there are three and two objects near RX J1257.5+2412 and RX J1239.3+2431, respectively. Low-resolution spectra of the optical objects nearest to the X-ray positions of RX J1239.3+2431, RX J1257.5+2412 and RX J1225.7+2055 were obtained on March 28, 1995 with the double spectrograph at the Palomar 200-inch Hale telescope (Tab. 6). Gratings with 316 and 300 lines/mm, resulting in a dispersion of 204 Å/mm and 140 Å/mm, were mounted in the red and the blue arm of the spectrograph. The dichroic separated the two sides at 5200 Å. The spectra covered 3500–5200 Å and 5200–7500 Å at a FWHM resolution of 3 Å and 6 Å, respectively. The spectra were corrected for bias and flatfield, and were wavelength calibrated using standard IRAF procedures. HD 84937 was used as standard for the flux calibration.

On June 29, 1995, a spectrum of the object closest to RX J1250.2+1923 was taken with COSMIC in the longslit spectrograph mode at the Palomar 200-inch Hale telescope (Tab. 6). A grism with 300 lines/mm, yielding a dispersion of 130 Å/mm, was used. The response of the spectrograph was derived from an observation of HZ 44, but no flux calibration was possible.

The optical positions of the identified AGN (given in Tab. 1) were determined using the APM finding chart programme which is based on the digitized Palomar Obser-vatory Sky Survey. Except object RX J1250.2+1923 the spectra were also used to derive an optical brightness (Tab. 2).

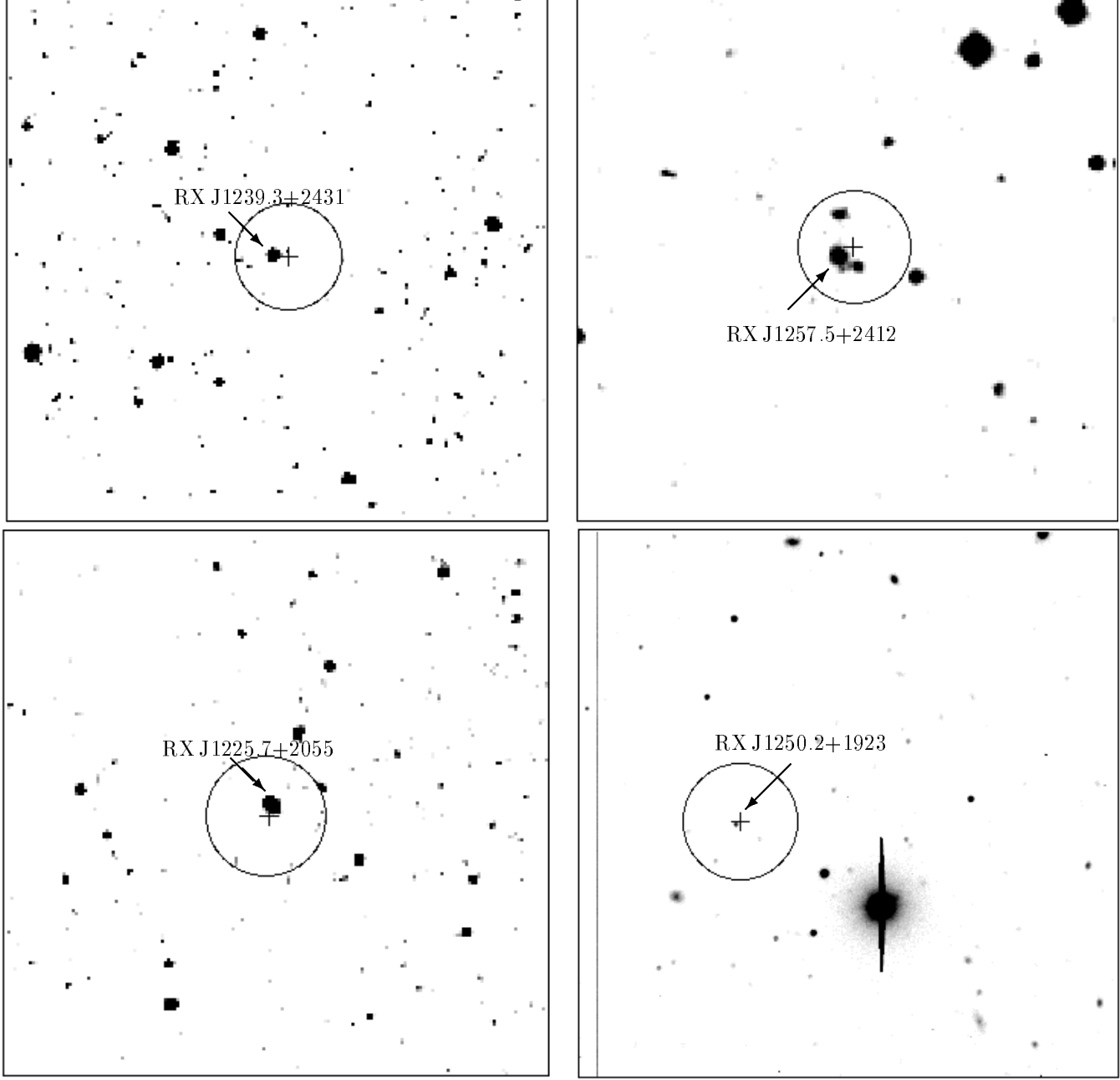

**Fig. 1.** Digitized Palomar Observatory Sky Survey charts for 3 of the 4 new AGN (indicated by arrows) and a 30 sec unfiltered image of RX J1250.2+1923 (bottom right). Crosses denote the best fit X-ray position while circles mark the $2\sigma$ X-ray error box (33″ radius). All charts are 5′ by 5′ with North at the top and East to the left.

### 2.2.1. RX J1239.3+2431

The spectrum of the central object at the location of RX J1239.3+2431 shows redshifted Balmer lines of hydrogen on top of a blue continuum. The strongest line is identified with H$\beta$ (see Fig. 2 for more identifications), which is consistent with the positions of the other strong lines detected. This results in a redshift of z=0.186. The high state of ionization, as indicated by the strength of [OIII]$\lambda$5007 relative to [OII]$\lambda$3727 (the latter is not detected in the spectrum), points to the Seyfert nature of the object while the faint absolute visual magnitude excludes a quasar identification. In particular, the clear presence of FeII emission and the fact that [OIII] < H$\beta$ (both unexpected for Seyfert 2 galaxies, which do not show FeII (e.g. Osterbrock 1989) and are characterized by an intensity ratio of [OIII]/H$\beta$ > 3 (Osterbrock and Shuder 1982)) lead to the classification of either a Seyfert 1 or Narrow Line Seyfert 1 (NLSy1 hereafter). The NLSy1 interpretation according to the classification of Osterbrock and Pogge (1985) is favoured by the weakness/absence of further narrow emission lines and particularly the not much broader width of H$\beta$ as compared to [OIII] (as quantified in section 2.3.).

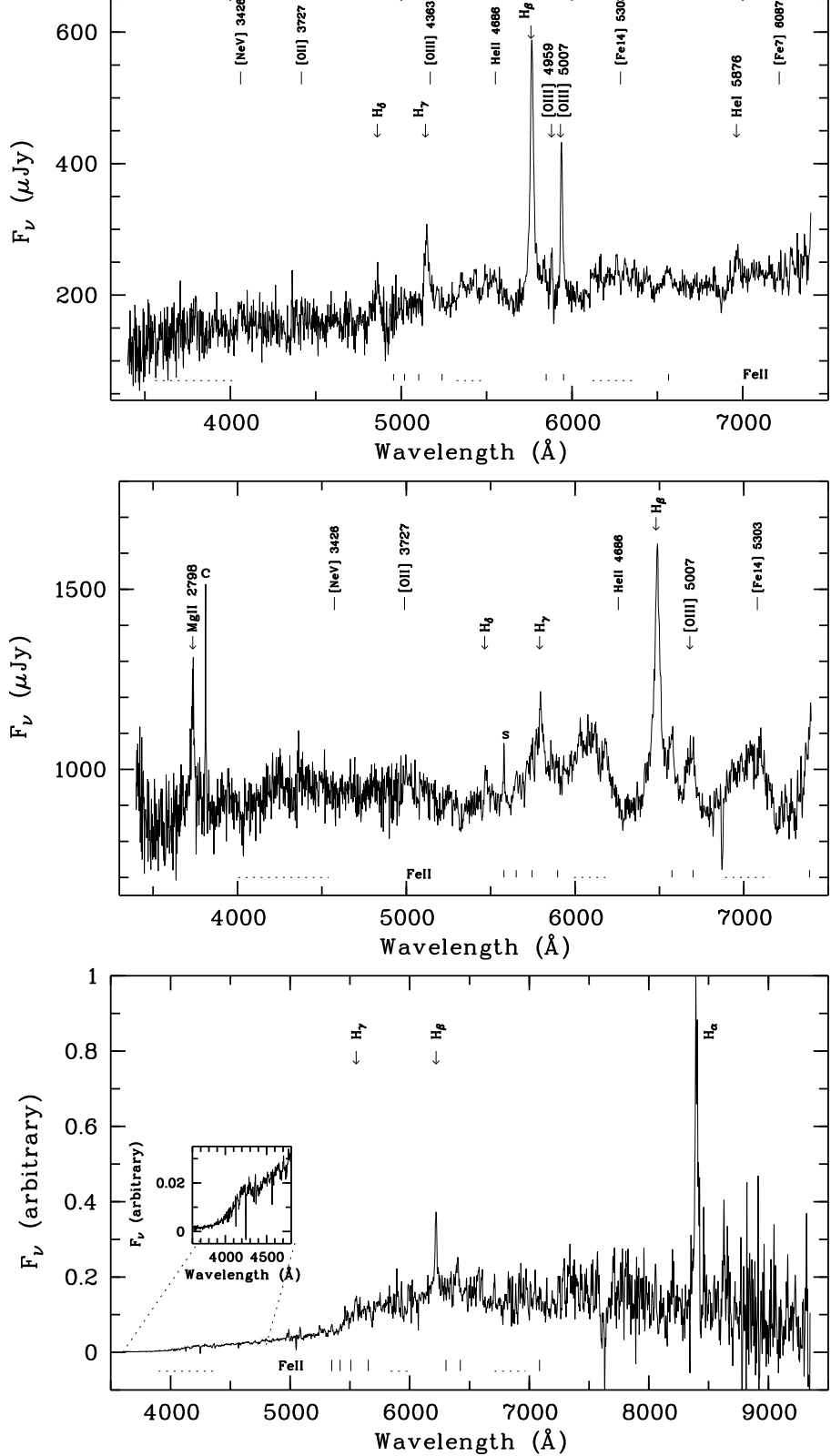

**Fig. 2.** Low-dispersion spectra of the new narrow line Seyfert 1s RX J1239.3+2431 (top panel), RX J1225.7+2055 (middle) and RX J1250.2+1923 (bottom). Some of the identified lines are marked (arrows), as well as narrow lines which might be expected but are not detected (vertical lines). At the bottom of each panel, the positions of some FeII lines (short vertical lines) and FeII line complexes (horizontal dotted lines) are indicated.

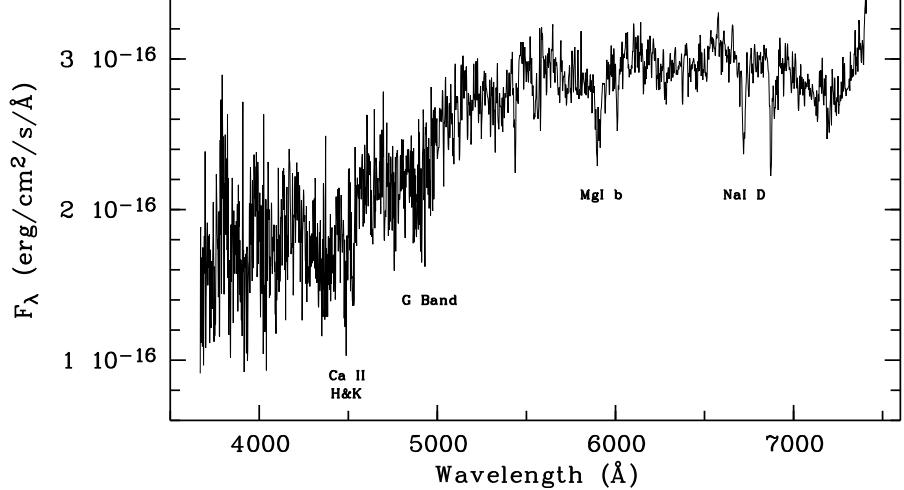

**Fig. 3.** Low-dispersion spectrum of the BL Lac object RX J1257.5+2412.

**Table 2.** Intensities and distances of the new AGN

| ROSAT Name | X-ray intensity (cts/sec) | Optical brightness from spectra[1] $m_B$ (mag) | plates[2] $m_{pg}$ (mag) | Radio flux at (GHz) 1.4 | 4.8 (mJy) | 8.4 | z | Luminosity distance (Mpc)[3] | $M_V$ (mag) |
|---|---|---|---|---|---|---|---|---|---|
| RX J1239.3+2431 | 0.26±0.1 | 18.2 | 16.0–17.3 | <0.8 | – | <0.15 | 0.186 | 1163 | −21.6 |
| RX J1257.5+2412 | 0.93±0.2 | 18.6 | 17.0–18.0 | 98 | 7.4[4] | 5.8 | 0.140 | 867 | −21.8 |
| RX J1225.7+2055 | 0.32±0.1 | 16.3 | 16.0–17.1 | <0.8 | – | <0.15 | 0.335 | 2153 | −24.1 |
| RX J1250.2+1923 | 0.18±0.05 | – | 20.5[5] | – | – | – | 0.280 | 1782 | ≈−23 |

[1] Intensity around 4350 Å.
[2] Maximum and minimum brightness on archival photographic plates of Sonneberg Observatory.
[3] Using $H_o=50$ km/s/Mpc and $q_o=0.5$
[4] This measurement is from the January 1992 observation, while all other radio observations are from August 1994.
[5] Brightness estimated on the Palomar Observatory Sky Survey blue print.

There is an additional object of 22 mag at 27″ distance to the X-ray position (east to the Seyfert galaxy). The noisy spectrum is blue and featureless, suggesting an identification as a faint blue galaxy which is not related to the X-ray source (based on the implied high $L_x/L_{opt}$ ratio).

### 2.2.2. RX J1257.5+2412

There are three objects within the RX J1257.5+2412 X-ray error circle. The northern and more distant object has a featureless spectrum pointing to a galaxy. The faintest of the three objects (south-west of the best fit X-ray position) is found to be an F or G star.

The optical spectrum of the third, brightest object is featureless without emission lines. Although some strong features from the underlying galaxy are visible the relative flux depression bluewards the Ca H/K break is less than 25%. Therefore this spectrum fulfills all spectroscopic BL Lac characteristics. Additional support for this interpretation comes from the relative fluxes in the radio, optical and X-ray bands (see below). Using the MgI b $\lambda$5176 absorption line we derive a redshift of z = 0.140. No correction for the presence of starlight has been made for the tabulated optical magnitudes and flux ratios.

**Table 3.** Measured and derived parameters of the H$\beta$ lines

|  | RX J1239.3+2431 | RX J1225.7+2055 | | | RX J1250.2+1923 |
|---|---|---|---|---|---|
|  | 1 component fit | 1 component fit | 2 component fit | | 1 component fit |
|  |  |  | broad | narrow |  |
| FWHM[1] [Å] | 19.8 | 24.7 | 52.9 | 12.4 | 13.8 |
| FWHM [km/s] | 1220 | 1530 | 3260 | 760 | 850 |
| $F_{H\beta}$ [$10^{-14}$ erg/cm$^2$/s] | 0.75±0.05[2] | 1.5 | 2.1 | 0.35 | – |
| $L_{H\beta}$ [$10^{42}$ erg/s] | 1.21±0.08[2] | 8.1 | 11.7 | 2.0 | – |

[1] Transformed into the AGN's rest frame.
[2] Errors due to uncertainty in fitting the continuum excluding the error in the absolute flux calibration.

### 2.2.3. RX J1225.7+2055

The spectrum of RX J1225.7+2055 exhibits strong emission lines at 3735 and 6489 Å on a strong blue continuum. Interpreting the 6489 Å line with H$\beta$ is consistent with all other lines identified (see Fig. 2) and leads to z=0.335. Only weak [OIII]$\lambda$5007 is present and [OII]$\lambda$3727 is missing. The FeII emission complexes in the spectrum reveal its Seyfert 1 nature. Their unusual strength suggests a NLSy1 classification, as is confirmed by the relatively small FWHM of H$\beta$ determined in section 2.3. Though the absolute visual magnitude is close to the (anyway arbitrarily defined) lower bound of quasars we will prefer calling the object a Seyfert galaxy.

### 2.2.4. RX J1250.2+1923

The spectrum of the faint object at the location of RX J1250.2+1923 again shows redshifted Balmer lines of hydrogen. The wavelength difference between these lines suggest the 8397 Å line to be H$\alpha$ which in turn leads to z=0.280. The emission feature longwards of H$\beta$ is thought to stem from FeII, as does the broad complex around 4100–4300 Å. With the weak or absent [OIII] emission and the absolute visual luminosity we identify RX J1250.2+1923 as NLSy1 galaxy.

### 2.3. Line widths and luminosities

In order to derive the FWHM of H$\beta$, the continuum was detemined over a broad wavelength region because of the presence of the many blended Fe-multipletts near H$\beta$. FWHMs were then deduced by fitting Gaussians to the lines. The instrumental effect on the measured line width was estimated to be less than 6%, and thus was not corrected for. Tab. 3 lists the derived line-profile parameters. The criterion for inclusion in the class of NLSy1, a H$\beta$ FWHM < 2000 km/s (Osterbrock and Pogge 1985, Goodrich 1989), is fulfilled by all three NLSy1s, RX J1239.3+2431, RX J1225.7+2055 and RX J1250.2+1923, thus confirming our former classification. It is surprising that all three X-ray selected objects are classified as NLSy1 according to their optical spectra.

Due to the missing absolute flux-calibration of the RX J1250.2+1923 spectrum no line parameters were derived except for the FWHM of H$\beta$. The H$\beta$ line parameters of the other objects were converted to luminosities assuming $H_o$ = 50 km/s/Mpc and $q_o$ = 0.5 (Tab. 3).

RX J1225.7+2055 shows evidence for a second, broader H$\beta$ component as compared to a single Gaussian which only would fit the narrow core of the line. However, the quality of the data does not allow a detailed multicomponent fit, which also strongly depends on the adopted continuum level and FeII contribution. Therefore, the 2-component fit presented in Tab. 3 only serves to determine an upper limit on the H$\beta$ luminosity. Additionally, H$\beta$ shows a slight blue-asymmetry in both, RX J1239.3+2431 and RX J1225.7+2055.

The spectrum of RX J1239.3+2431 exhibits strong enough [OIII] to allow the determination of its FWHM, leading to 710 km/s.

The amplitude ratio [OIII]/H$\beta$ in the core of the lines is $\approx 0.6$ for RX J1239.3+2431, < 0.25 for RX J1225.7+2055 and < 0.1 for RX J1250.2+1923.

### 2.4. Optical variability

We have examined the long-term optical behaviour of all our objects on photographic plates of the Sonneberg astrographs 400/1600 mm and 400/2000 mm. About 320 plates of the interval 1962–1995 of the field 26 Comae were used (Tab. 6). The limiting blue magnitude of the best plates is about $18^m$–$18^m5$ mag. In addition, for the object RX J1225.7+2055, 300 plates of the overlapping field 5 Comae, and for the objects RX J1257.5+2412 and RX J1250.2+1923 some hundred plates of the overlapping field 35 Comae could be used. The photometric calibration was done using several digitized photographic plates. The magnitudes were linked to the B magnitudes of a UBV sequence of several dozens of stars in the Comae region (Argue 1963) using a method of brightness determination on photographic plates developed by Kroll and Neuge-bauer (1993). For stars fainter than $17^m$ the sequence had to be extrapolated. Therefore, the given optical magnitudes have an internal error of about 0.1 mag and might be shifted absolutely by several tenth of a magnitude. The detailed measurements are available in electronic form at the CDS.

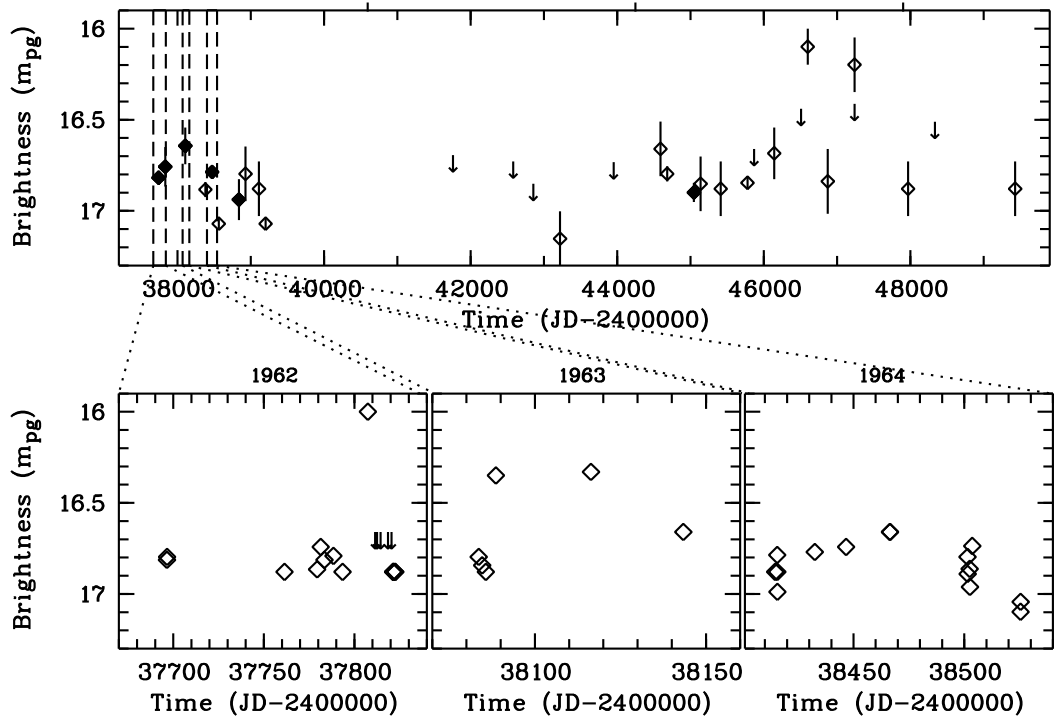

**Fig. 4.** Long-term lightcurve of RX J1239.3+2431 as deduced from the photographic field patrol of Sonneberg Observatory. Shown are three months averages (top panel) with filled (open) symbols denoting averages with more (less) than 5 individual measurements. Some upper limits are shown by arrows. The lower panel shows the individual measurements of the first three seasons with evidence of considerable intensity jumps within a few days (e.g. a rise by 0.5 mag within three days in the middle panel) and bright states of several weeks duration.

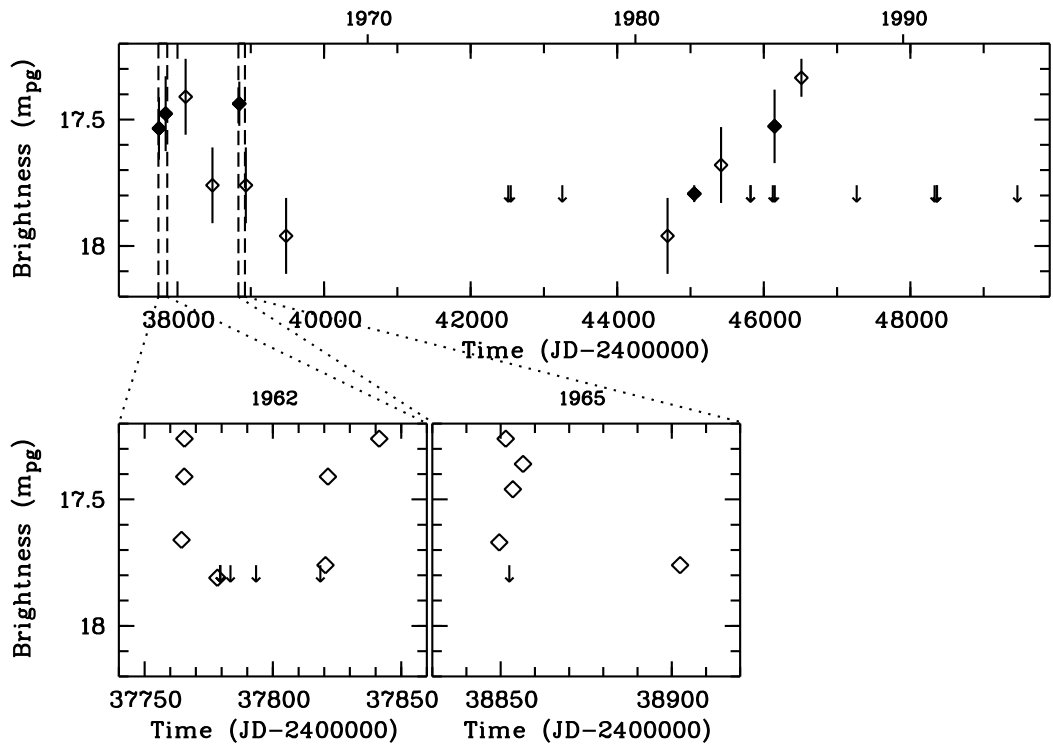

**Fig. 5.** Same for RX J1257.5+2412 except that filled (open) symbols denote averages with more (less) than 3 individual measurements.

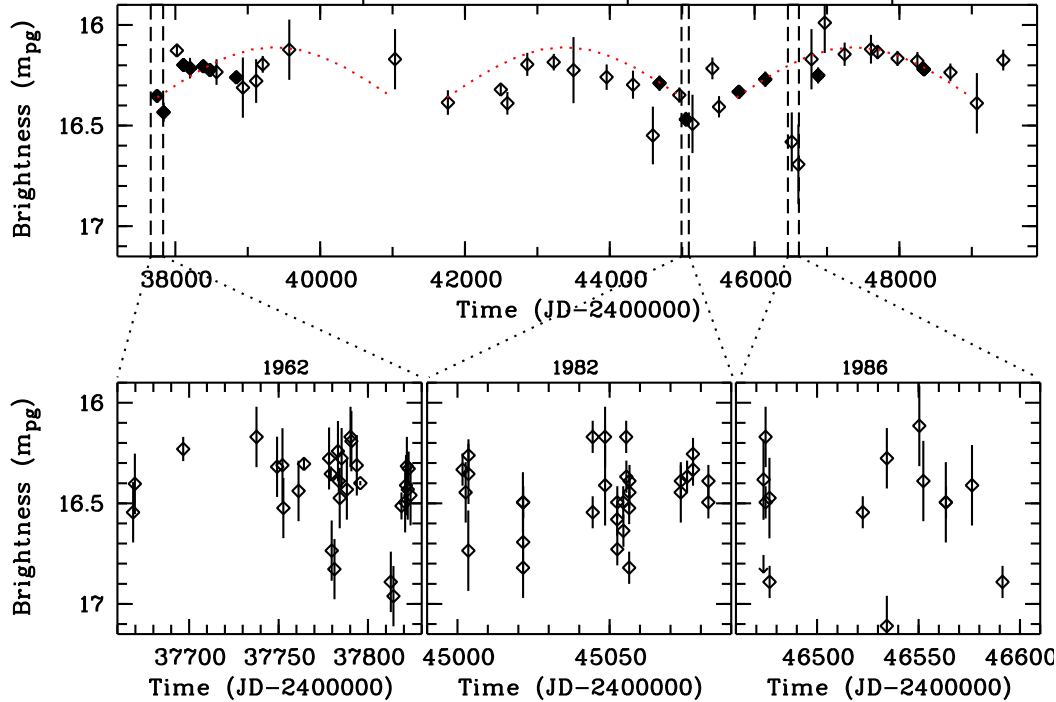

**Fig. 6.** Same for RX J1225.7+2055 except that filled (open) symbols denote averages with more (less) than 10 individual measurements. The time between the minima of the long waves (upper panel) is of the order of 11 years.

RX J1250.2+1923 proved to be too faint to be visible on Sonneberg plates. – The remaining objects turned out to be variable. In the fourth column of Tab. 1 the designation of a Sonneberg variable is given. The light variations can be described as follows:

RX J1239.3+2431 = S 10940 is visible only on good plates; in most cases its brightness is below the plate limit. The light curve is typical for Seyfert galaxies. It shows lively brightness changes: some spikes with a duration of several days or weeks can be seen (Fig. 4). The fastest variation with the largest amplitude in our data is a 0.55 mag jump within three days. After correction for the time stretching this timescale corresponds to a maximal size of the emission region of $\approx 6.5 \times 10^{15}$ cm.

RX J1257.5+2412 = S 10941 is clearly found to be variable, which besides the spectral characteristics is a classical property of BL Lac objects, the prototype of which was discovered by Hoffmeister (1929). Since S 10941 is mostly below the plate limit, no details on the form of the variations can be given except the rather slow fall of the mean brightness during the late 60ies and the slow rise in the early 80ies (Fig. 5). Variations on shorter timescales are not excluded, and indeed such variations are readily detected during seasons with better temporal coverage (lower panel of Fig. 5). Again, intensity drops and rises within a few days are measured with amplitudes up to nearly 1 mag.

The lightcurve of RX J1225.7+2055 = S 10942 exhibits long waves (several hundred to thousand days) superimposed on shorter (several tens of days) waves of small amplitude (Fig. 6). The large dispersion of the single observations indicates that there should be variations on a still shorter time scale. Indeed, there seem to be changes of several tenths of a magnitude, mostly minima, within a day. But the object being a Seyfert galaxy, the existence of short-term eclipsing light variations is improbable. CCD photometry performed in May 1995 on three night for 2–3 hours each failed to detect such short-term variations. Examples of the photographically detected variations are shown for two well covered seasons near the minima of the long-term waves and during a very deep minimum in 1986 (lower panel of Fig. 6). The largest observed variability timescale is 11 yrs (or 8 yrs in the Seyfert's rest frame), has an amplitude of 0.3 mag and is seemingly periodic (indicated by the dotted line in the upper panel of Fig. 6).

### 2.5. X-ray spectra

For the X-ray spectral analysis the source photons collected during the all-sky-survey scans were extracted with a radius of $5'$. The background was chosen at the same ecliptic longitude at $\approx 1°$ distance, corresponding to background photons collected typically 15 sec before or after the time of the source photons. Standard corrections were applied using the dedicated EXSAS software package (Zimmermann et al. 1994).

**Table 4.** X-ray spectral fit results of the new AGN[1]

|  | RX J1239.3+2431 | RX J1257.5+2412 | RX J1225.7+2055 | RX J1250.2+1923 |
|---|---|---|---|---|
| galactic $N_H$ ($10^{20}$ cm$^{-2}$)[2] | 1.26 | 1.34 | 2.86 | 2.07 |
| **power law** | | | | |
| photon index | $-4.3\pm1.3$ | $-2.4\pm0.2$ | $-3.6\pm1.1$ | $-5.4\pm2.0$ |
| $N_H$ ($10^{20}$ cm$^{-2}$) | 3.2 | 2.1 | 2.6 | 5.9 |
| Norm$_{powl}$ (ph/cm$^2$/s/keV)[3] | $2.4\times10^{-4}$ | $2.2\times10^{-3}$ | $5.7\times10^{-4}$ | $2.6\times10^{-4}$ |
| $\chi^2_{red}$ | 1.2 | 0.9 | 1.9 | 1.0 |
| **blackbody** | | | | |
| kT (eV) | $60\pm20$ | $210\pm20$ | $130\pm20$ | $65\pm20$ |
| $N_H$ ($10^{20}$ cm$^{-2}$) | 0.8 | 0.001 | 0.67 | 5.8 |
| Norm$_{bb}$ (ph/cm$^2$/s) | $6.3\times10^{-3}$ | $7.4\times10^{-3}$ | $5.9\times10^{-3}$ | 0.1 |
| $\chi^2_{red}$ | 1.4 | 2.8 | 2.2 | 1.1 |
| **disk blackbody plus power law**[4] | | | | |
| accretion rate ($\dot{M}_{Edd}$[5]) | 6.5 | – | 83.4 | 1.7 |
| $N_H$ ($10^{20}$ cm$^{-2}$) | 4.5 | – | 2.0 | 11.8 |
| Norm$_{db}$ (ph/cm$^2$/s/keV$^3$) | $3.8\times10^{-10}$ | – | $4.5\times10^{-12}$ | $1.1\times10^{-7}$ |
| Norm$_{powl}$ (ph/cm$^2$/s/keV) | $2.1\times10^{-4}$ | – | $3.1\times10^{-4}$ | $2.2\times10^{-4}$ |
| $\chi^2_{red}$ | 1.3 | – | 2.1 | 0.9 |
| effective temperature (eV) | 60 | – | 120 | 45 |
| powl component/total (%) | 4 | – | 25 | 0.1 |
| $L_x$ [0.1–2.4 keV] ($10^{45}$ erg/s) | 0.2–3.6 | 0.5–1.5 | 1.4–5.3 | 7.0–300 |
| $\alpha_{OX}$[6] | 1.17 | 0.67 | 1.51 | $\approx 0.8$ |
| $\alpha_{RO}$[7] | $<0.13$ | 0.59 | $<-0.03$ | – |

[1] All fits were performed with the redshift fixed at the optically determined value.
[2] Dickey & Lockman (1990)
[3] Normalized at 1 keV.
[4] The power law photon index was fixed at $-1.9$, and the mass of the central object at $10^6\ M_\odot$.
[5] $\dot{M}_{Edd}$ is defined as $L_{Edd}/(\eta\times c^2)$, where $\eta=1/12$ always; i.e. $\dot{M}_{Edd} = 2.66\times10^{-8}\ M/[M_\odot]\ M_\odot/\mathrm{yr} = 1.67\times10^{18}\ M/[M_\odot]$ g/s (newtonian).
[6] Ratio of the optical flux at 2500 Å in the AGN rest frame, and the X-ray flux at 2 keV: $\alpha_{OX} = -\log(S_{2keV}/S_{2500A})\ /\ 2.605$ according to Tananbaum et al. (1979).
[7] Ratio of the radio flux at 5 GHz and the optical flux: $\alpha_{RO} = \log(S_{5GHz}/S_{2500A})\ /\ 5.38$ according to Stocke et al. (1985).

According to the selection criteria all sources are dominated by emission below 0.5 keV though there is always weak, but non-zero emission above 0.5 keV. Motivated by the extreme hardness ratios we have tried some spectral fitting of the background-subtracted and vignetting-corrected source photons. We caution, however, that except for RX J1257.5+2412 all sources have only a relatively small number of photons (see column 2 in Tab. 1), and therefore all fitting results come with large statistical errors.

As a first step, we have fitted a power law model to the data. In all cases, this gives acceptable fits, i.e. in no case a second spectral component is necessary. The resulting photon indices range from $-2.4$ up to $-5.4$, and except for RX J1225.7+2055 all power law fits seem to require a larger absorbing column than the galactic hydrogen column. These and the remaining fit parameters are given in Tab. 4 which lists for all four new objects the total galactic absorption in the direction of the source (first row), the results of spectral fits of a power law, blackbody as well as disk blackbody plus power law model (for the case of the latter model including the effective temperature as well as the fractional luminosity of the power law component), and in the last rows the unabsorbed X-ray luminosity in the *ROSAT* band, and the optical to X-ray and radio to optical flux ratios.

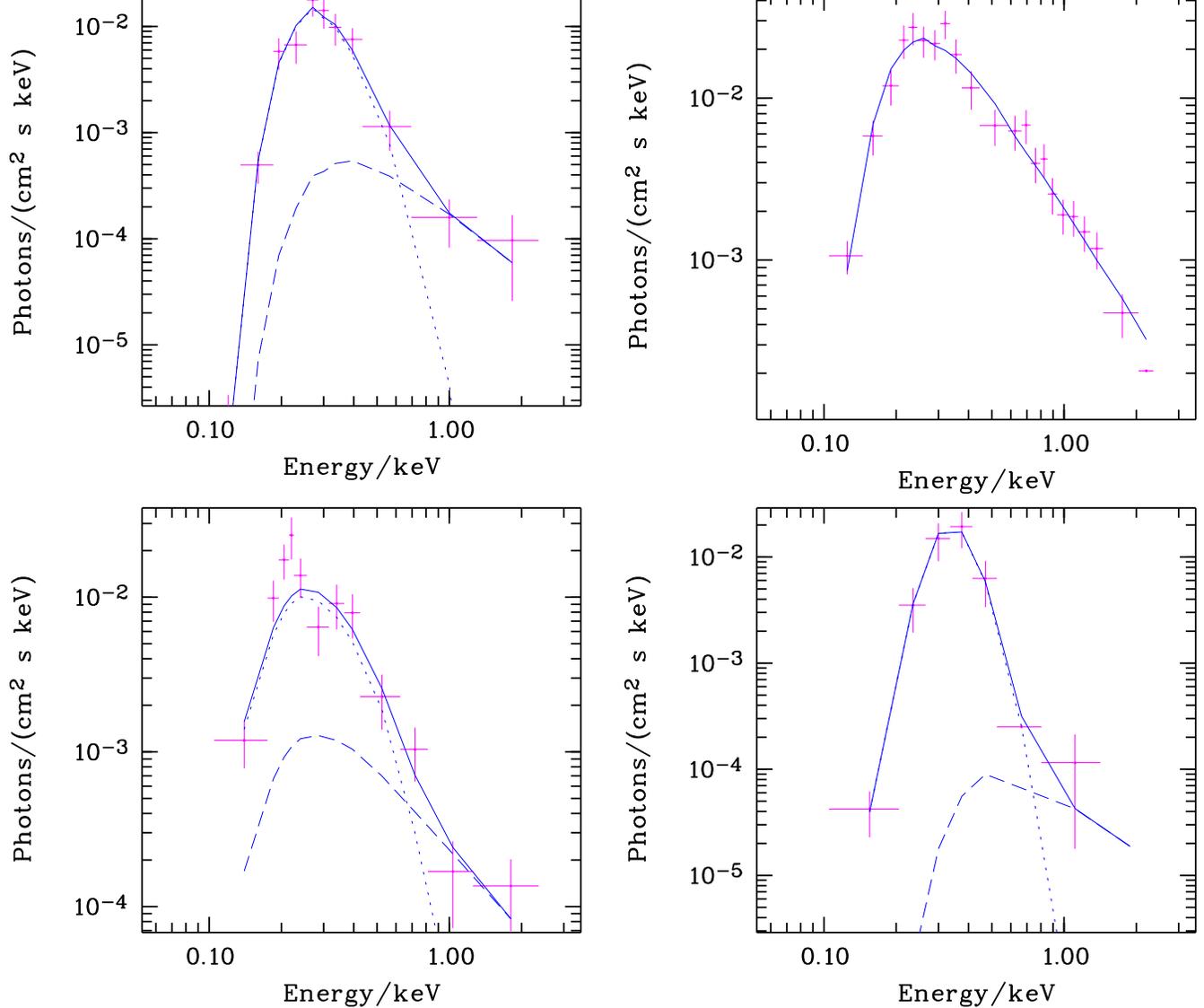

**Fig. 7.** Example of the spectral fits of RX J1239.3+2431 (top left), RX J1257.5+2412 (top right), RX J1225.7+2055 (bottom left) and RX J1250.2+1923. The BL Lac object was fitted with a single power law model (top right), while the 3 NLSy1s were fitted with a disk blackbody (dotted line) plus power law model (dashed line). For details see text and Tab. 4.

As a next step, fitting of a blackbody model (in the systems' rest frame) in all cases gives a considerably poorer reduced $\chi^2$ due to the fact that the hard energy tail has to be ignored. The same holds when fitting the standard disk blackbody model (Shakura and Sunyaev 1973). It is interesting to note that RX J1257.5+2412 requires practically no absorption at all when using a soft component above a power law. This strongly argues in favour of a single power law for the X-ray spectrum of this BL Lac object since it obviously cannot avoid the galactic absorption. Therefore, we have fitted for comparison purposes a disk blackbody plus a power law model (this combination is referred to as the accretion disk model in the following) to the X-ray data of the three NLSy1s. Since the slope of the power law is not constrained by the data we fixed the photon index at −1.9. Changing this fixed photon index to −1.5 has no effect on the best fit parameters except a < 5–10% change in the normalization.

Finally, we also applied a warm absorber model to the NLSy1 data (Tab. 5). The warm material was modeled using the photoionization code *Cloudy* (Ferland 1993). The ionizing spectral energy distribution (SED) incident on the clouds was assumed to originate from a point-like central energy source. Solar abundances were adopted (Grevesse and Anders 1989). As SED we have chosen a mean Seyfert continuum after Padovani and Rafanelli (1988) from the radio to the optical region with a break at 10$\mu$ and an energy index $\alpha=-2.5$ $\lambda$-longwards, an UV-EUV power law with $\alpha_{uv-x}=-1.4$ (Kinney et al. 1991) extending up to 0.1 keV, and an intrinsic X-ray power law with $\alpha=-0.9$ (Nandra and Pounds 1994) extending up to 100 keV followed by a break into the gamma-ray region. The index $\alpha_{uv-x}$ was later modified for RX J1239.3+2431 (to $\alpha_{uv-x}=-0.75$) and RX J1225.7+2055 (to $\alpha_{uv-x}=-1.9$) to account for the estimated EUV SED in these objects (see section 3.2.2 and Tab. 5). We note, however, that for the final best fit the ionization structure of the warm gas is dominated by the X-ray regime of the SED.

**Table 5.** Warm absorber fit results of the new NLSy1s[1]

| ROSAT Name | log U | log $N_w$ | log Norm[2] | $\alpha_{uv-x}$ | $\chi^2_{red}$ |
|---|---|---|---|---|---|
| RX J1239.3+2431 | $-0.1^{+0.3}_{-0.3}$ | $22.8^{+0.4}_{-0.3}$ | $-4.79^{+0.04}_{-0.05}$ | $-0.75$ | 1.0 |
| RX J1225.7+2055 | $0.8^{+0.3}_{-0.2}$ | $23.2^{+0.3}_{-0.3}$ | $-4.42^{+0.07}_{-0.07}$ | $-1.9$ | 1.8 |
| RX J1250.2+1923 | $0.4^{+0.3}_{-0.2}$ | $23.2^{+0.3}_{-0.4}$ | $-4.74^{+0.06}_{-0.06}$ | $-1.4$ | 0.9 |

[1] $N_H$ was fixed at its galactic value.
[2] In ph/cm$^2$/s/keV at 10 keV.

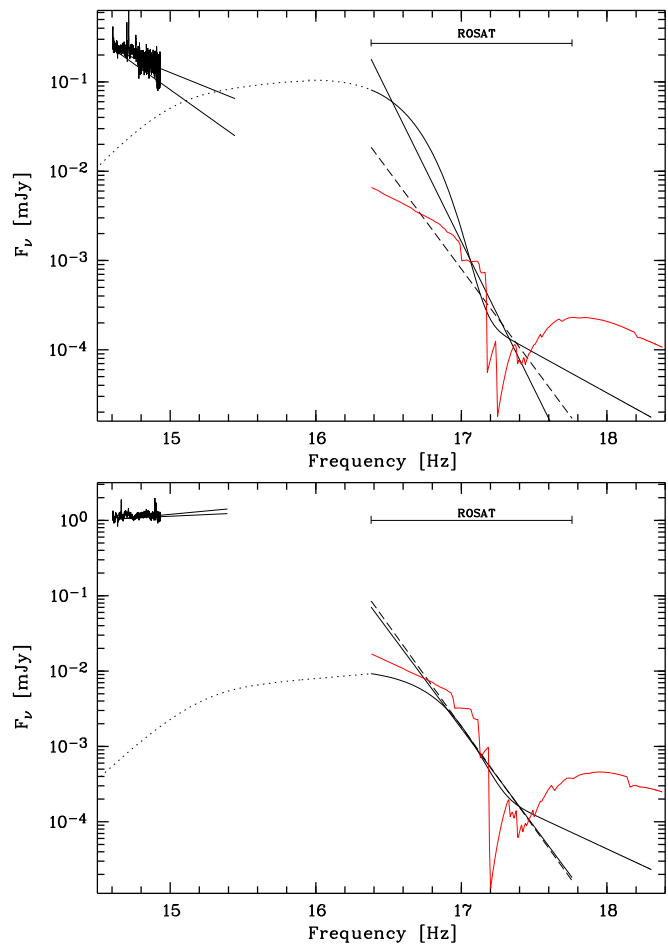

**Fig. 8.** Spectral energy distribution from the optical to soft X-rays in the NLSy1 RX J1239.3+2431 (top) and RX J1225.7+2055 (bottom). The solid lines in the right part are the (absorption corrected) best fit models applied to the *ROSAT* data. The dashed line corresponds to a power law model with the absorption fixed at the galactic $N_H$ value. The range of the low-energy ends of these curves represents the error in the flux determination at 0.1 keV. The solid lines drawn through the (extinction corrected) optical spectrum up to the Lyman limit visualize the lower and upper limit in the determination of the flux at the Lyman limit. The dotted line is the low-energy extension of the standard disk black body model which is simply added to show the general turn-over. It should be noted that the largest amplitude of the observed optical variability in RX J1225.7+2055 is a factor of 3, which cannot explain the rather large $\alpha_{OX}$.

We calculated a sequence of models with varying warm hydrogen column density $N_w$ and ionization parameter U, defined as $U=Q/(4\pi r^2 n_H c)$, where Q is the number rate of photons above the Lyman limit, r is the distance between nucleus and warm absorber, $n_H$ is the hydrogen density (fixed to $10^{9.5}$ cm$^{-3}$; this value is also used for the estimates in the discussion, but all derived quantities and in particular the X-ray absorption structure depend only weakly on $n_H$) and c the speed of light. Initially, the cold column density was left as an additional free parameter but turned out to always approach the galactic value and thus was fixed at that value for the final parameter estimates. The warm absorber fits were done in the rest frame of the Seyfert galaxies.

We find that a warm-absorbed flat power law provides a successful fit to the observations as well. As expected, the data do not allow to overcome the degeneracy between different combinations of U and $N_w$ that produce a similar ionization structure, i.e. several U-$N_w$ pairs represent the observed spectrum with comparable success. Instead of statistical errors we supply the range in both parameters according to a $\triangle \chi^2_{red} = 0.2$ (Tab. 5).

The multifrequency SEDs of RX J1239.3+2431 and RX J1225.7+2055 resulting from the combined optical and X-ray data are shown in Fig. 8.

### 2.6. X-ray intensity and luminosity

The *ROSAT* all-sky-survey observations (Tab. 6) consist of 31 scans with 9–30 sec duration each and spaced by the orbital period of the satellite of 96 min (or multiples). Due to these short exposure times and countrates of the sources between 0.2–1 cts/sec (Tab. 2) we have binned all photons of each scan into one time bin. The resulting lightcurves of all four objects are shown in Fig. 9. There are no drastic X-ray intensity variations over the two days of scanning observations. Whether or not the variations by a factor of two are real is hard to evaluate due to statistical uncertainties.

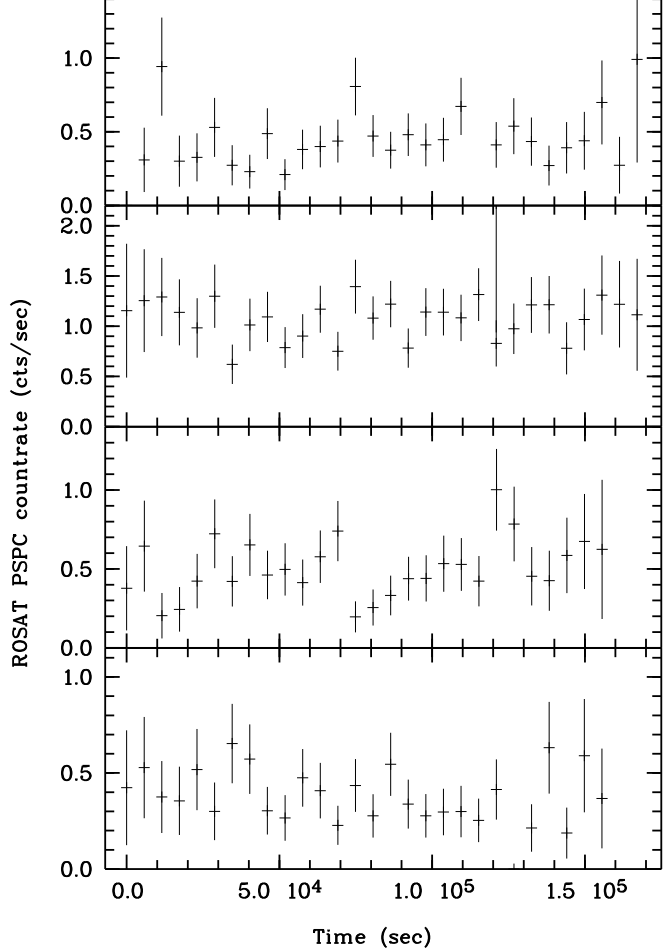

**Fig. 9.** X-ray survey lightcurves of all four objects RX J1239.3+2431, RX J1257.5+2412, RX J1225.7+2055 and RX J1250.2+1923 (from top to bottom). Each data point corresponds to the mean countrate during one scan with 9–30 sec length each.

The object RX J1257.5+2412 = 1ES 1255+244 was already "observed" in X-rays during 8 Einstein slews, resulting in the detection of 14 counts during the 34.2 sec of slew exposure (Elvis et al. 1992). Using Fig. 10.21 in the ROSAT AO1 document which gives the PSPC/IPC count rate ratio in dependence of the power law slope and the absorbing column we convert the background subtracted Einstein IPC rate of 0.36 cts/s with the best fit power law model parameters of Tab. 4 into an expected ROSAT PSPC rate of 1.01 cts/sec. The comparison with our ROSAT all-sky-survey measurement of 0.93 cts/sec demonstrates no variability within the errors of the measurement.

ison assumed the continuation of the steep X-ray slope as measured with the PSPC into the Einstein band which might be questionable. A hardening of the intrinsic spectrum above the ROSAT band would reduce the expected PSPC rate, but since only a part of the Einstein band would be affected, the conclusion of no X-ray variability is rather robust.

Converting the mean ROSAT all-sky-survey countrates into luminosities strongly depends on the spectral model adopted. We have used the accretion disk model for the three NLSy1s and the power law model for RX J1257.5+2412, and give as "errors" of these estimates for the NLSy1s the range of luminosities resulting from applying those of the different spectral models of Tab. 4 which yield acceptable fits.

Since we have no contemporaneous optical measurements to the X-ray survey observations, and all but one source show optical variability by a factor of 2–4, the optical luminosities during December 1990 are uncertain also. Thus, the derived $\alpha_{OX}$ ratios for our objects (Tab. 4) can only serve as order of magnitude estimates.

### 2.7. Radio Observations

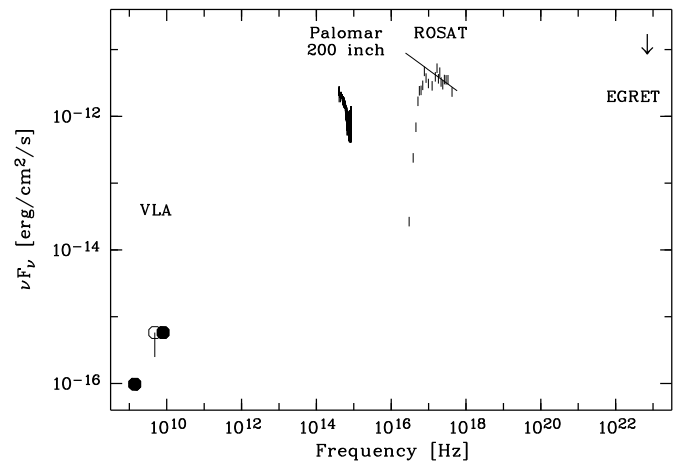

**Fig. 10.** Broad band energy distribution of the new BL Lac source RX J1257.5+2412 as determined from our non-contemporaneous data. Filled circles denote VLA measurements from 1994, the open circle that from 1992 (the vertical line visualizes the amplitude of variation within these two years). The arrow gives the 90% upper limit above 100 MeV from the EGRET phase 1–3 database (Maddox 1995). The ROSAT X-ray spectrum is shown as observed with the best fit power law model added (extinction corrected) while the optical data are extinction corrected.

All sources except RX J1250.2+1923 were mapped in August 1994 with the Very Large Array (VLA) radio telescope in its hybrid BC configuration at 1.4 and 8.4 GHz for all three objects at the two frequencies. Due to radio interference, the noise limit at 1.4 GHz stayed well above the theoretical limit of 190 µJy.

In the field of RX J1257.5+2412 an unresolved radio source was detected at both radio frequencies (see Table 2). The radio position is R.A. $(2000.0) = 12^h 57^m 31\overset{s}{.}9$ and Decl. $(2000.0) = +24°12'40''$ corresponding to a $2\overset{''}{.}5$ offset from the optical APM-chart position $(12^h 57^m 31\overset{s}{.}8 \; +24°12'38'')$. However, since the APM position is derived from a blended object, its position might have a larger than usual error. We therefore adopt the radio position as the more precise one (see Tab. 1). The offset from the nominal ROSAT all-sky-survey position is $8''$.

We also analyzed an observation taken from the VLA archive (Schachter 1995) which was performed in January 1992 at 4.8 GHz, and included the radio flux from the again unresolved source in Table 2. The flux of this 1992 observation is about a factor of 8 lower than the interpolation of the flux measurements in 1994 between 1.4 GHz and 8.4 GHz. Thus, besides the optical brightness changes the BL Lac object RX J1257.5+2412 is also strongly variable at radio frequencies.

## 3. Discussion

### 3.1. The BL Lac object RX J1257.5+2412

Recent investigations of X-ray spectra of BL Lac's (Ciliegi et al. 1995) have shown that their average photon index in the soft X-ray range (0.2–2.4 keV) is −2.23±0.17 for X-ray selected and −2.52±0.73 for radio selected objects. RX J1257.5+2412 with its photon index of −2.4 is among the softest X-ray selected BL Lac's. This is interesting because the unified model of AGN with X-ray selected BL Lac's viewed in average at higher angles between viewing and jet axis than radio selected BL Lac's predicts that there should be a difference between the X-ray spectra of the unbeamed (or at least wide opening angle), X-ray selected and beamed, radio selected BL Lac's. With a decreasing angle the inverse compton emission becomes more important due to Doppler boosting, thus dominating at high energies with respect to synchrotron emission and flattening the X-ray spectra. Therefore, the discovery of RX J1257.5+2412 argues against this clear cut between radio and X-ray selected BL Lac's.

Its multifrequency colours of $\alpha_{OX} = 0.67$ and $\alpha_{RO} = 0.59$ clearly show the X-ray dominated SED of this object (see also Fig. 10) for which often the abbreviation XBL is used (Stocke et al. 1985). Since the identification of serendipitously found Einstein sources it has been realized that BL Lac objects are luminous X-ray emitters with $L_x = 10^{43}...10^{46}$ erg/s in the 0.3–3.5 keV band. Among the brightest extragalactic X-ray sources BL Lac objects form a high portion. Nevertheless, bright XBLs like RX J1257.5+2412 (with a corresponding X-ray flux in the Einstein band of $F_X = 8 \times 10^{-12}$ ergs/cm$^2$/s) are rare. Published surface densities of X-ray selected BL Lac objects from Einstein (Maccacaro et al. 1989) and preliminary results from the ROSAT all-sky-survey (Nass et al. 1995) lead to the expectation of less than one object with such a flux in a 100 square degree area. However, it is not possible to verify these earlier results with this one object.

**Table 6.** Observation log of all four sources

| X-ray Observations | | |
| --- | --- | --- |
| Date | exposure | off-axis |
| Dec. 10–17, 1990 | 470 sec | 0–55' |
| Optical observations[1] | | |
| Date | Type | Observatory |
| 1962–1995 | photographic patrol | Sonneberg |
| May 1995 | CCD photometry | Sonneberg |
| March 28, 1995 | spectroscopy | Palomar |
| June 29, 1995 | spectroscopy | Palomar |
| VLA Radio observations[2] | | |
| Date | Mode | Frequency |
| Jan. 31, 1992 | snapshot | 4.8 GHz |
| Aug. 9, 1994 | snapshot | 1.4 GHz |
| Aug. 9, 1994 | snapshot | 8.4 GHz |

[1] Photographic patrol observations and spectroscopy in March 1995 of objects RX J1239.3+2431, RX J1257.5+2412 and RX J1225.7+2055; CCD photometry in May 1995 of object RX J1225.7+2055; spectroscopy in June 1995 of object RX J1250.2+1923.

[2] Object RX J1257.5+2412 in January 1992, and objects RX J1239.3+2431, RX J1257.5+2412 and RX J1225.7+2055 in August 1994.

### 3.2. The Narrow Line Seyfert 1 objects

#### 3.2.1. Implications from the optical variability

The optical variability of Seyfert 1 galaxies is usually slow and irregular with small amplitudes. The typical timescales of variations are months to years while faster variations are rare. The extremes in both, amplitude and timescale, are characterized as "some Seyfert nuclei have been observed to vary up to two magnitudes within a few years, while variations of several tenths of a magnitude can occure within days or weeks" (Hamilton et al. 1978). From this point of view our two variable Seyfert galax-extraordinary objects.

Studies of the optical long-term behaviour of AGN have revealed indications of periodic or quasi-periodic fluctuations for some of them. Rest frame "periods" of the order of ten years have been reported for the quasars 1217+02, 1004+13, 2349−01 (Pica et al. 1980), 3C 120 and 3C 345 (Webb et al. 1988), and 0736+017 (Wallinder et al. 1992). However, in all cases no definite conclusions were possible mainly due to the fact that the available time base was not much longer than these long periods. Our database over more than 30 years covers three cycles of RX J1225.7+2055 (Fig. 6), thus supplying one of the strongest evidences for periodic or quasi-periodic variations in AGN.

Several possibilities have been discussed to explain such long-term periodic or quasi-periodic variations (see e.g. Wallinder et al. 1992 for a review): (1) A bright spot in the accretion disk with a lifetime of several orbital periods might be eclipsed by the outer part of the disk, (2) Precessing jets in near pole-on geometry or shocks propagating along the jet, (3) Dwarf-novae type disk instabilities with matter accumulation times much longer than the consumption phase at high luminosity, (4) Thermal limit-cycle disk oscillations in the inner part of the disk. In application to our Seyfert galaxies the jet models seem to be ruled out because of the (generally believed) small contribution of the jet to the overall luminosity. If there would be a substantial part of the optical radiation escaping to us without impinging on the broad line region (BLR), this would cause additional problems (for instance for the $L_{H\beta}$ production).

The rest frame time scale of 8 yrs corresponds to the Kepler frequency at a distance of $6\times10^{15}$ ($6\times10^{16}$) cm of a $10^6$ ($10^9$) $M_\odot$ Schwarzschild black hole. At these distances the temperature is still too high for dust formation, so that absorption at this location can be excluded. On the other hand, these distances are very similar to the ones derived from the light travel time argument using the shortest observed optical variations. This suggests a common emission process being responsible for both, short- and long-term variations. All proposed scenarios for long-term variability have problems with such a combination.

In the case of RX J1225.7+2055 there seems to be an additional problem with accretion disk models because the multifrequency energy distribution is hard to reconcile with these kind of models (see below and Fig. 8). This suggest either that the optical emission does not stem from the disk and hence all the above mentioned models would not be applicable, or that the X-ray part of the spectrum does not correspond to the Wien tail of the disk.

### 3.2.2. Implications from the optical spectra

Several comparisons of X-ray and optical (line and continuum) properties of AGN have revealed various correlations ratio with steeper X-ray spectra, and a decreasing $L_{[OIII]}$ with increasing $L_{FeII}$ (Puchnarewicz et al. 1992, Boroson and Green 1992, Bade 1993, Laor et al. 1994, Boller et al. 1995). All our three objects fit in these trends though they are selected by their X-ray properties (extremely soft X-ray spectra). In particular, RX J1250.2+1923 with its very steep X-ray spectrum has the smallest H$\beta$ FWHM, representing the most extreme object so far in the FWHM$_{H\beta}$–$\Gamma_x$ distribution of NLSy1s (cf. Fig. 8 in Boller et al. 1995). Previous investigations have found an increasing scatter of the X-ray continuum slope with decreasing FWHM (Bade 1993, Boller et al. 1995).

Among our objects, the spectrum of RX J1225.7+2055 exhibits particularly strong FeII complexes, about a factor 10 more than in RX J1239.3+2431. Integrating over the FeII $\lambda$4570 Å complex in RX J1225.7+2055 yields F = 2.6–3.1$\times 10^{-14}$ erg/cm$^2$/s which again ranges among the highest FeII fluxes among similar distant (z) NLSy1s (Puchnarewicz et al. 1992). Dividing by the H$\beta$ flux derived from the 1 component fit (as the lower limit of the H$\beta$ luminosity) results in a FeII/H$\beta$ ratio of 1.75–2.15. This is not unusually high despite the high FeII flux. This ratio is dropped even more if we use the H$\beta$ flux derived from the two component fit (see Tab. 3). Thus, the FeII/H$\beta$ ratio does not necessarily represent a good measure of the FeII flux.

In a slightly different approach of the same problem, Boller et al. (1995) have argued that the unusually large ratio of FeII/H$\beta$ observed in many NLSy1 might be explained by weaker than usual H$\beta$ emission. The two objects for which we derive H$\beta$ luminosities do not behave according to this hypothesis. While the H$\beta$ luminosity of RX J1239.3+2431 just corresponds to the mean value of the Sy1 sample in Padovani and Rafanelli (1988), that of RX J1225.7+2055 (and even the lower limit derived from the one component fit) is at the high end of this distribution.

$L_{H\beta}$ allows to estimate the minimal number of hydrogen-ionizing photons Q isotropically emitted by the central continuum source. Assuming T=10.000 K and using Tabs. 2.1 and 4.2 of Osterbrock (1989) results in Q = 2.1$\times 10^{12}$ $L_{H\beta}$. Using the one-component fits to H$\beta$ this translates to log Q = 54.40 for RX J1239.3+2431 and log Q = 55.23 for RX J1225.7+2055.

The three X-ray spectral models used for the fitting and discussed in the next section show different flux distributions in the EUV spectral region. However, all these models have to fulfill the constraint to provide enough photons to account for the observed H$\beta$ luminosity. In the following extrapolations, we always consider only the simplest case for each of these models, i.e. the accretion disk model without a possible additional underlying UV-X power law, or similarly a single power law or the warm absorber model without a possible additional EUV bump component. In order to estimate a lower limit on Q for served optical-UV power law to the Lyman limit with the steepest slope allowed by the data. For the pure power law and warm absorber models we then constructed (and integrated over) a power law from the Lyman limit to the low-energy end of the ROSAT spectrum (0.1 keV) using the best fit X-ray parameters and applying the absorption correction. In case of the accretion disk model the predicted EUV flux distribution was directly integrated. The resulting $\alpha_{uv-x}$ for RX J1239.3+2431 and RX J1225.7+2055 are given in Tab. 5. The calculations were done in the objects' restframes.

We find that each model can account for the observed H$\beta$ luminosity, except for RX J1225.7+2055, for which the accretion disk model results only in $Q \approx Q_{L_{H\beta}}$. This relation would imply a covering factor of unity for the BLR gas, thus blocking our view to the X-ray source which is in contradiction to what is observed. This low Q value results from the fact, that the predicted EUV bump does not match the extrapolated observed flux at the Lyman limit but falls short by two orders of magnitude.

Under the assumption that Q has a value between the above deduced lower limit and an upper limit estimated with the flattest possible optical-UV power law (with no additional bump in the unobserved EUV region) we discuss the resulting properties of the warm absorber in item 3 of section 3.2.3.

3.2.3. Implications from the X-ray spectral fits

Three distinctly different spectral distributions are possible to explain the X-ray data. Either these systems have single power law spectra extending with the same slope towards higher energies or they have (blackbody-like) excess emission on top of a standard Seyfert power law spectrum. Alternatively, a steep spectrum in the soft X-ray region can result from warm absorption of an intrinsically flat spectrum. While the X-ray data alone are not sufficient to decide between these possibilities in our cases, it might be instructive to explore them in some more detail.

1. *Single steep power law:* As stated above, single power law spectra fit our X-ray data well. With the resulting best-fit photon indices these new sources belong to the steep (soft) end of their population known so far.
   As a cautious note we point out that the energy coverage together with the limited energy resolution of the ROSAT PSPC does not allow to unambiguously constrain multicomponent spectral models (except possibly for very high signal-to-noise ratio observations of a few bright objects). This often serves, quite correctly, as justification for using a single power law model for the spectral fitting of ROSAT AGN data. However, the finding of a steep spectrum (high photon index) does not imply that the X-ray spectrum continues to be steep outside the ROSAT band, especially above 2.5 keV. Therefore, high photon indices derived from simplistic description of a phenomenon which can be any of the above mentioned three different spectral distributions. Thus, measuring a broader spectral range is essential to reveal the true energy distribution.
   While for RX J1225.7+2055 the slope of the X-ray power law can be easily imagined to be extended through the UV and smoothly matching the optical spectral slope, the situation is quite different for RX J1239.3+2431 (see Fig. 8). Here, the X-ray intensity at 0.1 keV is similar to (or even exceeds) the extrapolated intensity at the Lyman limit. Thus, one would have to invoke a completely flat spectrum between these two energies which in turn would imply two unnatural breaks. We therefore conclude that the single steep power law is an unsatisfactory model for the combined X-ray and optical data.

2. *Soft excess on flat power law:* Given the observational evidence of the presence of a blue/UV bump, a more natural interpretation of the X-ray spectra than steep power laws would assume an additional soft component on top of a power law of typical slope (photon index of −1.9). Since this soft component often is related with the emission of an optically thin accretion disk (e.g. Czerny and Elvis (1987), Madau (1988)), we have used a disk blackbody model as soft component. Corresponding to the low statistics we have fixed the mass of the central object thus fitting only the accretion rate and the normalization. We have performed several fits with the (fixed) mass between $10^4$ and $10^9$ $M_\odot$ and verified that (1) the ratio between accretion rate and mass (the maximum temperature of the disk) is always the same and (2) that the intensity in the power law component does not change. The maximum temperature of the disk as given in Tab. 4 turns out to be very similar to the temperatures of single blackbody fits.
   The extrapolation of the disk blackbody spectrum in its standard form (Shakura and Sunyaev 1973) towards the UV and optical range is problematic due to several effects (see extensive discussion in Greiner et al. 1994). We only note here that the observed optical flux is higher by a factor 10–500 than the extrapolation of the best fit models for all systems.

3. *Flat power law with warm absorption:* Another possibility to produce a steep spectrum in the soft X-ray region is via warm absorbing material along the line of sight to the central continuum source. In such warm gas (T of the order of $10^5$ K) highly ionized metal ions imprint ionization edges onto the soft X-ray continuum passing the gas. These edges have been seen in the spectra of several Seyfert galaxies. One expects to also see objects with even stronger warm absorption, i.e. deep edges which recover only at hard energies above 2.5 keV, leaving mainly the "down-turning" part visible in the ROSAT spectral band.

nario (as in other proposed models) it is not straightforward to explain the narrowness of the Balmer lines in NLSy1s. On the other hand, considering the scatter in the FWHM$_{H\beta}$-$\Gamma_x$ correlation, NLSy1s might well be a heterogeneous group with more than one mechanism at work to make them look different from 'normal' Sy1s.

The best-fit column densities, log $N_w \approx$ 22.8 – 23.2, are rather similar for the three objects, whereas the ionization parameters, log U $\approx$ –0.1 – 0.8 vary by nearly an order of magnitude. However, both come with large errors. The range in U indicates a higher state of ionization than is typically found in the 'usual' BLR, consistent with the warm absorbers observed so far. Taking the mean Q (as defined in section 2.5. and estimated from the multi-wavelength SED in section 3.2.2), i.e. $4.2 \times 10^{56}$ for RX J1225.7+2055 and $8.8 \times 10^{54}$ for RX J1239.3+2431, the absorber-intrinsic H$\beta$ emission for a thin shell geometry with full covering is calculated to be about $10^{41.89}$ for RX J1225.7+2055 and $10^{41.23}$ for RX J1239.3+2431. This corresponds to 1/10 and 1/7 of the observed $L_{H\beta}$, respectively. Scaling the predicted [Fe14]$\lambda$5303 emission of the warm material in order not to conflict with the observed upper limit (see Fig. 2) constrains the covering factor of the gas to $\leq$ 1/6 in RX J1239.3+2431 and is consistent with 1 in RX J1225.7+2055. Given the high discovery rate of supersoft AGN among X-ray selected ones (section 3.3), the covering factors indeed have to be high to account for this fact.

The two composite spectra including a flat power law model (i.e. above items 2 and 3) have two important advantages: The flat power law with the Seyfert-1 typical index $\Gamma_x$=–1.9, as used in our modelling, is consistent with the nonthermal pair models usually invoked to explain the X-ray spectrum in Seyferts, which naturally predict this spectral slope (e.g. Svensson 1994). Moreover, the favourite model for producing the FeII emission, i.e. via reprocessing of X-rays deep within the BLR clouds, could be reconciled much easier with a flat instead of a steep soft X-ray spectrum.

The major difference between accretion disk and warm absorber model, as far as the flux distribution is concerned, is demonstrated in Fig. 8. These different EUV SEDs would be expected to differently influence the BLR (and NLR) line emission (although both cannot easily explain the absence/weakness of the usual NLR lines, if they are assumed to illuminate an otherwise normal, i.e. Seyfert-like, NLR (Komossa and Greiner, 1995)). The second major difference is the different intensity of the hard X-ray flux at a few keV. The X-ray intensity of the warm absorber model is nearly one order of magnitude larger than that of the flat power law added to the disk blackbody component. This prediction should be tested with inate between these two models.

As a consequence, the higher hard X-ray continuum of the warm absorber model relative to an accretion disk continuum may influence the strength of the FeII emission. It is interesting to note that the ratio in the integrated intrinsic luminosities (for both, the warm absorber and the accretion disk model) between RX J1239.3+2431 and RX J1225.7+2055 of a factor of 10 is similar to the ratio of the integrated FeII ($\lambda$4570 Å complex) luminosities.

### 3.3. Source statistics

The 100 square degree field searched systematically in the *ROSAT* all-sky-survey data is located at a galactic latitude of bII = 77–88 degree. Correspondingly, more than half of the detected 238 X-ray sources are of extragalactic nature. The new soft AGN reported here are the softest extragalactic sources in this sample, thus representing the soft end of the AGN spectral distribution. Though the optical identification of this sample is not yet complete the following estimate can be made on the expected number of new supersoft AGN to be discovered in the *ROSAT* all-sky-survey data: The number of supersoft AGN according to the selection criteria given in section 2.1. divided by the searched area results in a number density of $\approx$0.04 per square degree. Excluding a 20 degree zone around the galactic equator one might expect on the order of 1500 supersoft AGN in the *ROSAT* survey which is about 7% of the number of expected AGN of all subclasses. Though the above numbers are small and admittedly uncertain, they are considered to be a lower limit, however, because this estimate of the number density of supersoft AGN is based only on sources with more than 80 photons (one of our selection criteria) while the expected number of the total *ROSAT* survey AGN population ($\sim$25000) includes all survey sources with more than typically 10 photons.

Taking into account the intensity cut in our selection, the above percentage of supersoft AGN readily increases. Out of the 238 X-ray sources in our field there are 20 sources with more than 80 counts which divide up into 8 stars and 12 extragalactic objects, mainly AGN. Among these, the herewithin reported four supersoft AGN are equally distributed according to X-ray intensity, i.e. the relative number density of supersoft AGN is independent of X-ray intensity. Any extrapolation to faint objects has to scope with two major selection effects: First, uneven absorption over the sky would affect the detection probability more severely at low galactic latitudes. Second, fainter (and thus more distant) objects are typically harder due to their higher redshift. Both these effects work against the detection of supersoft AGN, thus leaving the following estimates still lower limits. Thus, ignoring for the moment these biases and extrapolating this number density downwards to the all-sky-survey detection threshold would imply that basically every third X-ray selected AGN is super-

Having about 120 AGN among the 238 sources, the 12 AGN with more than 80 counts represent only 10%. With the above extrapolation we expect to have 40 instead of 4 supersoft AGN in our 100 square degree field. (We note that this field has a mean exposure time of about 450 sec, and thus the uneven exposure of the sky during the ROSAT all-sky-survey does not invalidate this extrapolation.) With these numbers the conclusion seems justified that supersoft X-ray spectra are common among X-ray selected AGN samples and do not only constitute the "soft tail" of the AGN spectral distribution.

## 4. Conclusions

We have discovered four AGN in the ROSAT all-sky-survey data with very steep X-ray spectra. We identify three of these objects as Narrow Line Seyfert 1 galaxies, and the fourth as BL Lac object. If one requires one model to explain the optical to X-ray energy distribution of all three NLSy1s then the warm absorber model is preferred. In this case, an additional EUV spectral component would have to be assumed in RX J1225.7+2055 while in the other two sources the extrapolation of the warm absorber model is consistent with the observed optical intensities. If different models are allowed for different sources then a single steep power law model seems to be the simplest explanation for the optical to X-ray energy distribution in RX J1225.7+2055, and a warm absorber model for RX J1239.3+2431.

The small–FWHM H$\beta$ lines, strong FeII emission and weak [OIII] emission in the three Narrow Line Seyfert 1 galaxies is in line with known correlations with respect to the steepness of the X-ray spectra in AGN. While one object shows particularly strong FeII emission (in terms of flux and luminosity), its FeII/H$\beta$ ratio is usual. This indicates that the FeII/H$\beta$ ratio is not a well suited observational indicator of the "FeII problem".

We have discovered strong optical variability in the BL Lac object and two of the Seyfert galaxies using photographic plates of the field patrol of Sonneberg Observatory (the third NLSY1 is too weak for a study on archival plates). All objects show strong short-term variability (few days or less). The long-term variation in RX J1225.7+2055 seems to be periodic or quasi-periodic with three cycles of 11 yrs covered by the data.

Extrapolating the number of supersoft AGN found in the 10 × 10 degree field under the assumption that they are equally distributed with respect to observable X-ray fluxes, we conclude that about 30% of all X-ray selected AGN could be supersoft. While we find three of the four objects to be NLSy1s, the distribution of supersoft AGN over the different sub-classes remains to be determined by optical follow-up studies of larger samples of ROSAT sources.

ing and calibrating the spectroscopic observations with the double spectrograph at Palomar Observatory. We are grateful to R. McMahon for installing the software used for the APM finding charts. We thank Gary Ferland for generously providing Cloudy. JG is supported by the Deutsche Agentur für Raumfahrtangelegenheiten (DARA) GmbH under contract No. FKZ 50 OR 9201, and GAR and PK by the Deutsches Elektronen-Synchrotron (DESY-PH) under contract 05 2SO524. RD gratefully acknowledges the continuing hospitality in the group of S.R. Kulkarni during his long term visit at the California Institute of Technology. NB is supported by DFG under Re353/22-3. The ROSAT project is supported by the German Bundesministerium für Bildung und Wissenschaft (BMBW/DARA) and the Max-Planck-Society. The National Radio Astronomy Observatory is operated by Associated Universities, Incorporated, under cooperative agreement with the National Science Foundation. Based on photographic data of the National Geographic Society – Palomar Observatory Sky Survey (NGS-POSS) obtained using the Oschin Telescope on Palomar Mountain. The NGS-POSS was funded by a grant from the National Geographic Society to the California Institute of Technology. The plates were processed into the present compressed digital form with their permission. The Digitized Sky Survey was produced at the Space Telescope Science Institute under US Government grant NAG W-2166.